\PassOptionsToPackage{hyphens}{url}
\PassOptionsToPackage{prologue,usenames,table,dvipsnames,x11names}{xcolor}
\documentclass[sigconf,screen,balance,natbib=false,authorversion]{acmart}

\usepackage{graphicx}
\usepackage{booktabs}
\usepackage{multirow}
\usepackage[abbreviations]{foreign}
\usepackage[inline]{enumitem}
\usepackage{hyperref}
\usepackage{cleveref}\usepackage{pifont}
\usepackage{subcaption}
\usepackage[subtle]{savetrees}
\usepackage[margin=0pt,skip=6pt,belowskip=0pt,font={small,stretch=0.9},labelfont=bf,justification=centering]{caption}
\usepackage{tabularx, makecell, booktabs}
\usepackage{tikz}
\usepackage[compact]{titlesec}
\usepackage{xcolor}
\usepackage[
  backend=biber,
  doi=false,
  isbn=false,
  style=ACM-Reference-Format,
]{biblatex}

\hypersetup{
  colorlinks = true, urlcolor   = blue, linkcolor  = blue, citecolor  = red   }

\DeclareFieldFormat{titlecase}{\MakeSentenceCase*{#1}}
\AtEveryBibitem{\clearfield{pages}\clearfield{volume}\clearlist{location}\clearfield{number}\clearfield{pagetotal}\clearfield{numpages}\ifentrytype{online}{}{\clearfield{url}}\ifentrytype{inproceedings}{\clearlist{publisher}\clearfield{articleno}}{}} 

\def\isExtendedCiteEnabled{1}

\ifnum\isExtendedCiteEnabled=1
  \newcommand{\extcite}[2][null]{~\cite{#2}}\else
  \newcommand{\extcite}[2][null]{\def\paramOne{#1}\def\paramNull{null}\ifx\paramOne\paramNull \else ~\cite{#1}\fi }
\fi

\newcolumntype{R}{>{\raggedleft\arraybackslash}X}
\newcolumntype{P}[1]{>{\raggedleft\arraybackslash}p{#1}}
\newcolumntype{q}[1]{>{\centering\arraybackslash\hspace{0pt}}p{#1}}

\definecolor{prioritycolor}{HTML}{969bce}
\newcommand*{\priority}[1]{\raisebox{-1pt}{\begin{tikzpicture}[scale=0.15]\draw (0,0) circle (1);
    \fill[fill opacity=1,fill=prioritycolor] (0,0) -- (90:1) arc (90:90-#1*3.6:1) -- cycle;
    \end{tikzpicture}}}

\newcommand{\compfull}{\priority{100}}
\newcommand{\comppart}{\priority{50}}
\newcommand{\compnone}{\priority{0}}

\AtBeginDocument{\providecommand\BibTeX{{\normalfont B\kern-0.5em{\scshape i\kern-0.25em b}\kern-0.8em\TeX}}}

\acmConference[SysTEX '22 Workshop]{the 5th Workshop on System Software for Trusted Execution}{@ASPLOS}{Lausanne, Switzerland}
\setcopyright{none}
\acmISBN{}
\acmDOI{10.48550/arXiv.2204.06790}

\makeatletter
\begin{document}

\title[An Exploratory Study of Attestation Mechanisms for Trusted Execution Environments]{An Exploratory Study of Attestation Mechanisms\\ for Trusted Execution Environments}

\author{Jämes Ménétrey}
\orcid{0000-0003-2470-2827}
\email{james.menetrey@unine.ch}
\affiliation{\institution{University of Neuchâtel}
  \country{Switzerland}
}

\author{Christian Göttel}
\orcid{0000-0002-4465-6197}
\email{christian.goettel@unine.ch}
\affiliation{\institution{University of Neuchâtel}
  \country{Switzerland}
}

\author{Marcelo Pasin}
\orcid{0000-0002-3064-5315}
\email{marcelo.pasin@unine.ch}
\affiliation{\institution{University of Neuchâtel}
  \country{Switzerland}
}

\author{Pascal Felber}
\orcid{0000-0003-1574-6721}
\email{pascal.felber@unine.ch}
\affiliation{\institution{University of Neuchâtel}
  \country{Switzerland}
}

\author{Valerio Schiavoni}
\orcid{0000-0003-1493-6603}
\email{valerio.schiavoni@unine.ch}
\affiliation{\institution{University of Neuchâtel}
  \country{Switzerland}
}

\begin{abstract}
Attestation is a fundamental building block to establish trust over software systems.
When used in conjunction with trusted execution environments, it guarantees that genuine code is executed even when facing strong attackers, paving the way for adoption in several sensitive application domains.
This paper reviews existing remote attestation principles and compares the functionalities of current trusted execution environments as Intel SGX, Arm TrustZone and AMD SEV, as well as emerging RISC-V solutions.
\end{abstract}


\maketitle

\sloppy

\begin{table*}[!t]
\centering
\smaller
\setlength{\tabcolsep}{4pt}
\rowcolors{1}{gray!10}{gray!0}
\begin{tabularx}{\textwidth}{X|q{16mm}|c|q{12mm}q{12mm}q{12mm}|cccc}
    \toprule
    \rowcolor{gray!25}
    & & & \multicolumn{3}{c|}{\textbf{SEV}} & \multicolumn{4}{c}{\textbf{RISC-V}} \\
    \rowcolor{gray!25}
    \multirow{-2}{*}{\textbf{Features}} & \multirow{-2}{*}{\textbf{SGX}} & \multirow{-2}{*}{\textbf{TrustZone}} & Vanilla & SEV-ES & SEV-SNP & Keystone & Sanctum & TIMBER-V & LIRA-V \\
    \midrule
    Integrity & \compfull & \compnone & \compnone & \compnone & \compfull & \compfull & \compnone & \compnone & \compnone \\
    Freshness & \compfull & \compnone & \compnone & \compnone & \compfull & \compfull & \compnone & \compnone & \compnone \\
    Encryption & \compfull & \compnone & \compfull & \compfull & \compfull & \compfull & \compnone & \compnone & \compnone \\
    Unlimited domains & \compfull & \compnone & \comppart & \compfull & \compfull & \compfull & \compfull & \compfull & \compnone \\
    Open source & \comppart & \comppart & \compnone & \compnone & \compnone & \compfull & \compfull & \compfull & \compnone \\
    Local attestation & \compfull & \compnone & \compnone & \compnone & \compnone & \compnone & \compfull & \compfull & \compnone \\
    Remote attestation & \compfull & \comppart & \compfull & \compfull & \compfull & \compfull & \compfull & \compfull & \compfull  \\
    API for attestation & \compfull & \comppart & \compnone & \compnone & \compfull & \compfull & \compfull & \compfull & \compfull \\
    Mutual attestation & \compnone & \comppart & \compnone & \compnone & \compnone & \compnone & \compnone & \compnone & \compfull \\
    User-mode support & \compfull & \compfull & \compfull & \compfull & \compfull & \compfull & \compfull & \compfull & \compnone \\
    Industrial TEE & \compfull & \compfull & \compfull & \compfull & \compfull & \compnone & \compnone & \compnone & \compnone \\
    &&&&&&&&& \\
    \rowcolor{gray!0}
    \multirow{-2}*{\makecell[l]{Isolation and\\attestation granularity}} & \multirow{-2}*{\makecell[c]{Intra-address\\space}} & \multirow{-2}*{\makecell[c]{Secure\\world}} & \multirow{-2}*{\makecell[c]{Virtual\\machine}} & \multirow{-2}*{\makecell[c]{Virtual\\machine}} & \multirow{-2}*{\makecell[c]{Virtual\\machine}} & \multirow{-2}*{\makecell[c]{Secure\\world}} & \multirow{-2}*{\makecell[c]{Intra-address\\space}} & \multirow{-2}*{\makecell[c]{Intra-address\\space}} & \multirow{-2}*{\makecell[c]{Intra-address\\space}}\\
    \rowcolor{gray!10}
    &&&&&&&&& \\
    \rowcolor{gray!10}
    &&&&&&&&& \\
    \rowcolor{gray!10}
    \multirow{-3}*{\makecell[l]{System support for\\isolation}} & \multirow{-3}*{\makecell[c]{Microcode}} & \multirow{-3}*{\makecell[c]{Secure\\monitor}} & \multirow{-3}*{\makecell[c]{Firmware}} & \multirow{-3}*{\makecell[c]{Firmware}} & \multirow{-3}*{\makecell[c]{Firmware}} & \multirow{-3}*{\makecell[c]{Secure\\monitor\\+ PMP}} & \multirow{-3}*{\makecell[c]{Secure\\monitor\\+ PMP}} & \multirow{-3}*{\makecell[c]{Tagged\\memory\\+ MPU}} & \multirow{-3}*{\makecell[c]{PMP}}\\
    \bottomrule
    \end{tabularx}
    \caption{\label{tab:features}Comparison of the state-of-the-art TEEs.}
\end{table*}

\section{Introduction}
Confidentiality and integrity are essential building blocks for secure computer systems, especially if the underlying system cannot be trusted.
For example, video broadcasting software can be tampered with by end-users who circumvent digital rights management.
Also, virtual machines are candidly open to the indiscretion of their cloud-based untrusted hosts.
The availability of Intel SGX, AMD SEV, RISC-V, Arm TrustZone-A/M \emph{Trusted Execution Environments} (TEEs) into commodity processors significantly helps to build trusted applications.
In a nutshell, TEEs execute software with stronger security guarantees, including privacy and integrity, without relying on a trustworthy operating system.

Remote attestation allows trusting a specific piece of software by verifying its authenticity and integrity.
Through remote attestation, one ensures to be communicating with a specific, trusted (attested) program remotely.
TEEs can support and strengthen the attestation process, ensuring the software being attested is shielded against powerful attacks and isolated from the outer system.
However, TEEs are used for attestation using a variety of different techniques. This survey reviews the current practices regarding \emph{remote attestation mechanisms} for TEEs~\cite{maene2017attestation}, covering a selection of TEEs of the four major architectures, namely Intel SGX, Arm TrustZone-A/-M, AMD SEV and a few emerging implementations for RISC-V. 

This paper is organised as follows.
In \S\ref{sec:att}, we describe the general principles of attestation and highlight the differences between local and remote attestation.
In \S\ref{sec:atttee} we survey the existing support for attestation in the TEE implementations currently available in commodity hardware.
We conclude in \S\ref{sec:conc} discussing some future directions.
 \section{Attestation}
\label{sec:att}

Attestation is an operation through which one software environment proves that a specific program is running in specific hardware.
Local attestation is used locally, between two software environments running in the same hardware, where one trusted environment proves its identity to another environment, hosted on the same system.
Attestation is based on a local hardware-bound secret called \emph{root of trust}, used to generate keys to sign the code being executed.
One can assess whether an attestation is genuine by verifying that the signature (of a specific processor) matches the code supposed to be in execution.
The result of an attestation can be used to establish new secrets, \ie to establish secure communication channels between both environments. 

Remote attestation can establish trust between software environments running in \emph{different} hardware.
This document adopts the terminology from IETF~\cite{ietf-rats-architecture-12}.
A \emph{relying party} wishes to establish a trusted relationship with an \emph{attester}, leveraging a \emph{verifier}.
The attester provides the state of its system, indicating the hardware and the software stack that runs on its device by collecting a set of \emph{claims}.
An example of a claim is the device's application code measurement, typically a cryptographic hash.
Claims are collected and cryptographically signed to form an \emph{evidence}, later asserted or denied by the verifier.
Once the attester is proven genuine, the relying party can safely interact with it and, for instance, transfer confidential data.

The problem of remotely attesting software has been extensively studied, and many implementations already exist based on software, hardware, or a combination of both.
Software-based remote attestation\extcite[seshadri2005pioneer]{seshadri2005pioneer,10.1007/978-3-540-74477-1_97,steiner2019towards} does not depend on any particular hardware, and it is adapted to low-cost devices.
Hardware-based remote attestation can rely on tamper-resistant hardware as, for instance, a \emph{Trusted Platform Module} (TPM) to ensure that the claims are trustworthy~\cite{6104065}, or a \emph{Physical Unclonable Function} (PUF) that prevents impersonations by using unique hardware marks produced at manufacture\extcite[6881436]{6881436,feng2018aaot}.
Other approaches exist, such as exposing a hardware secret fused in a die exclusively to a trusted environment.
Hybrid solutions combine hardware and software\extcite[10.1145/3460120.3484532]{6800458,carpent2018remote,10.1145/3460120.3484532,236230}, in an attempt to leverage advantages from both sides.
In \S\ref{sec:atttee} we describe how TEEs support remote attestation.

Trusted applications may need stronger trust assurances by ensuring both ends of a secure channel are attested.
For example, when retrieving confidential data from a sensing IoT device (for sensitive data), the device must authenticate the remote party, while the latter must ensure the sensing device has not been spoofed or tampered with.
\emph{Mutual attestation} protocols have been designed to appraise the trustworthiness of both end devices involved in a communication.
We report in \S\ref{sec:atttee} how mutual attestation has been studied in the context of TEE.
 \section{Issuing attestations using TEEs}
\label{sec:atttee}

Several solutions exist to implement hardware support for trusted computing, and TEEs are particularly promising.
Typically, a TEE consists of isolating critical components of the system, \eg, portions of the memory, denying access to more privileged but untrusted systems, such as kernel and machine modes.
Depending on the implementation, it guarantees the confidentiality and the integrity of the code and data of trusted applications, thanks to the assistance of CPU security features.
This work surveys modern and prevailing TEEs from processor designers and vendors with remote attestation capabilities for commodity or server-grade processors, namely Intel SGX~\cite{cryptoeprint:2016:086}, AMD SEV~\cite{amd-sev}, and Arm TrustZone~\cite{pinto2019demystifying}.
Besides, RISC-V, an open ISA with multiple open source core implementations, ratified the \emph{Physical Memory Protection} (PMP) instructions, offering similar capabilities to memory protection offered by aforementioned technologies~\cite{riscvratifications}.
As such, we also included many emerging academic and proprietary frameworks that capitalise on standard RISC-V primitives, which are Keystone~\cite{10.1145/3342195.3387532}, Sanctum~\cite{197162}, TIMBER-V~\cite{Weiser2019TIMBERVTM} and LIRA-V~\cite{9474324}.
Finally, among the many other technologies in the literature, we omitted the TEEs lacking remote attestation mechanisms (\eg IBM PEF~\cite{10.1145/3447786.3456243}) as well as the TEEs not supported on currently available CPUs (\eg Intel TDX~\cite{sardar2021demystifying}, Realm~\cite{armrealm} from Arm CCA~\cite{armcca}).

\hfill

\subsection{TEE cornerstone features}
We propose a series of cornerstone features of TEEs and remote attestation capabilities and compare many emerging and well-established state-of-the-art solutions in Table \ref{tab:features}.
Each feature is detailed below and can either be missing (\compnone), partially (\comppart) or fully (\compfull) available.
While we define these features below, we elaborate further about each TEE in the remainder of the section.

\emph{Integrity}: an active mechanism preventing DRAM of TEE instances from being tampered with.
\emph{Freshness}: protecting DRAM of TEE instances against replay and rollback attacks.
\emph{Encryption}: TEE instances' DRAM is encrypted for providing some assurance that no unauthorised access or memory snooping of the enclave occurs.
\emph{Unlimited domains}: many TEE instances can run concurrently, while the TEE boundaries (\eg isolation, integrity) between these instances are guaranteed by hardware. Partial fulfilment means that the number of domains is capped.
\emph{Open source}: indicate whether the solution is either partially or fully publicly available.
\emph{Local attestation}: a TEE instance can attest to another instance running on the same system.
\emph{Remote attestation}: a TEE instance can be attested by remote parties. Partial fulfilment means no built-in support, but extended by the literature.
\emph{API for attestation}: an API is available by the trusted applications to interact with the process of remote attestation. Partial fulfilment means no built-in support, but extended by the literature.
\emph{Mutual attestation}: the identity of the attestation and the verifier are authenticated upon remote attestations. Partial fulfilment means no built-in support, but extended by the literature.
\emph{User mode support}: state whether the trusted applications are hosted in user mode, according to the processor architecture.
\emph{Industrial TEE}: contrast the TEEs used in production and made by the industry from the research prototypes designed by the academia.
\emph{Isolation and attestation granularity}: the level of granularity where the TEE operates for providing isolation and attestation of the trusted software.
\emph{System support for isolation}: the hardware mechanisms used to isolate trusted applications.

\subsection{TEEs and remote attestation}

The attestation of software and hardware components require an environment to issue evidences securely.
In practice, this role is usually assigned to some mechanism that cannot be tampered with.
These environments rely on measuring the executed software (\eg, by hashing its code) and combining that output with cryptographical values derived from the hardware, such as a root of trust fused in the die or a physical unclonable function.
We analysed today's practices for the leading processor vendors for issuing cryptographically signed evidences.

\begin{figure}[!t]
    \centering
    \includegraphics[width=\columnwidth]{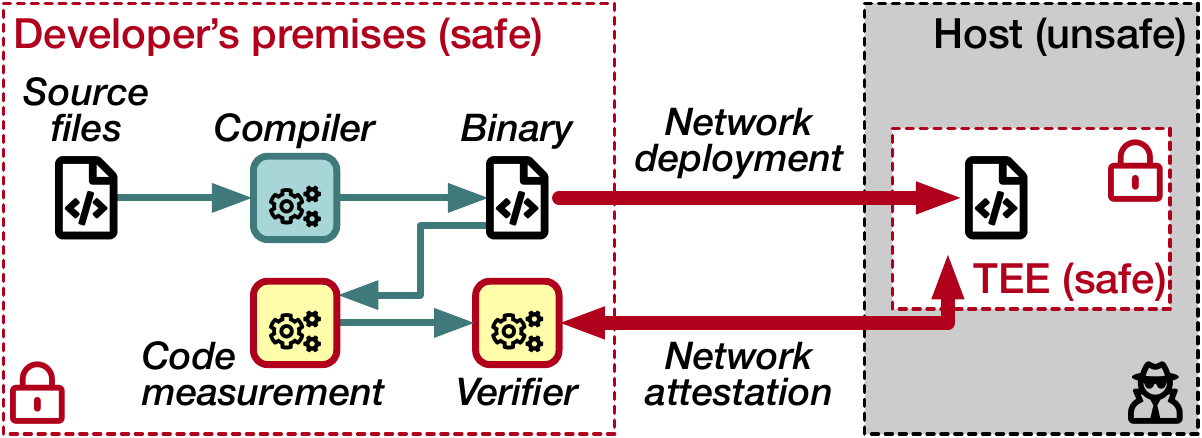}
    \caption{The workflow of deployment and attestation of TEEs.}
    \label{fig:workflow}
\end{figure}

Figure \ref{fig:workflow} illustrates the generic workflow for the deployment of trusted applications.
Initially, the application is compiled and measured on the developers' premises.
It is later transferred to an untrusted system, executed in the TEE.
The trusted application then communicates with a verifier to establish a trusted channel.
The TEE environment helps this transaction by exposing an evidence to the trusted application, which adds key material to it, preventing an attacker from eavesdropping on the communication.
The verifier asserts the evidence comparing it to a list of reference values to identify genuine instances of trusted applications.

\subsection{Intel SGX}
\label{sec:sgx}
\begin{figure*}[!t]
    \centering
    \begin{subfigure}[b]{0.3\textwidth}
        \centering
        \includegraphics[width=\textwidth]{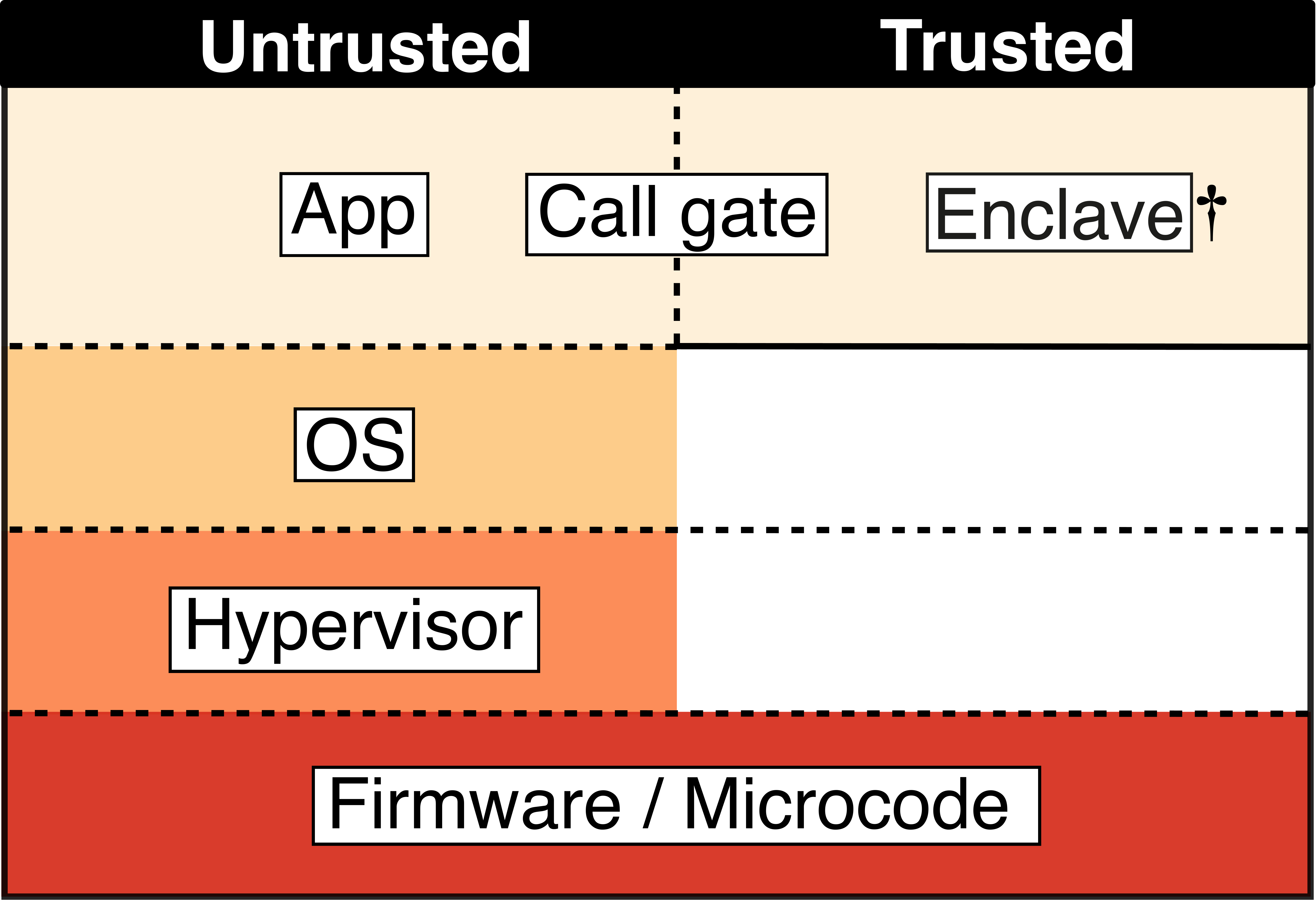}
        \caption{Intel SGX}
        \label{fig:sgx}
    \end{subfigure}
    \hfill
    \begin{subfigure}[b]{0.3\textwidth}
        \centering
        \includegraphics[width=\textwidth]{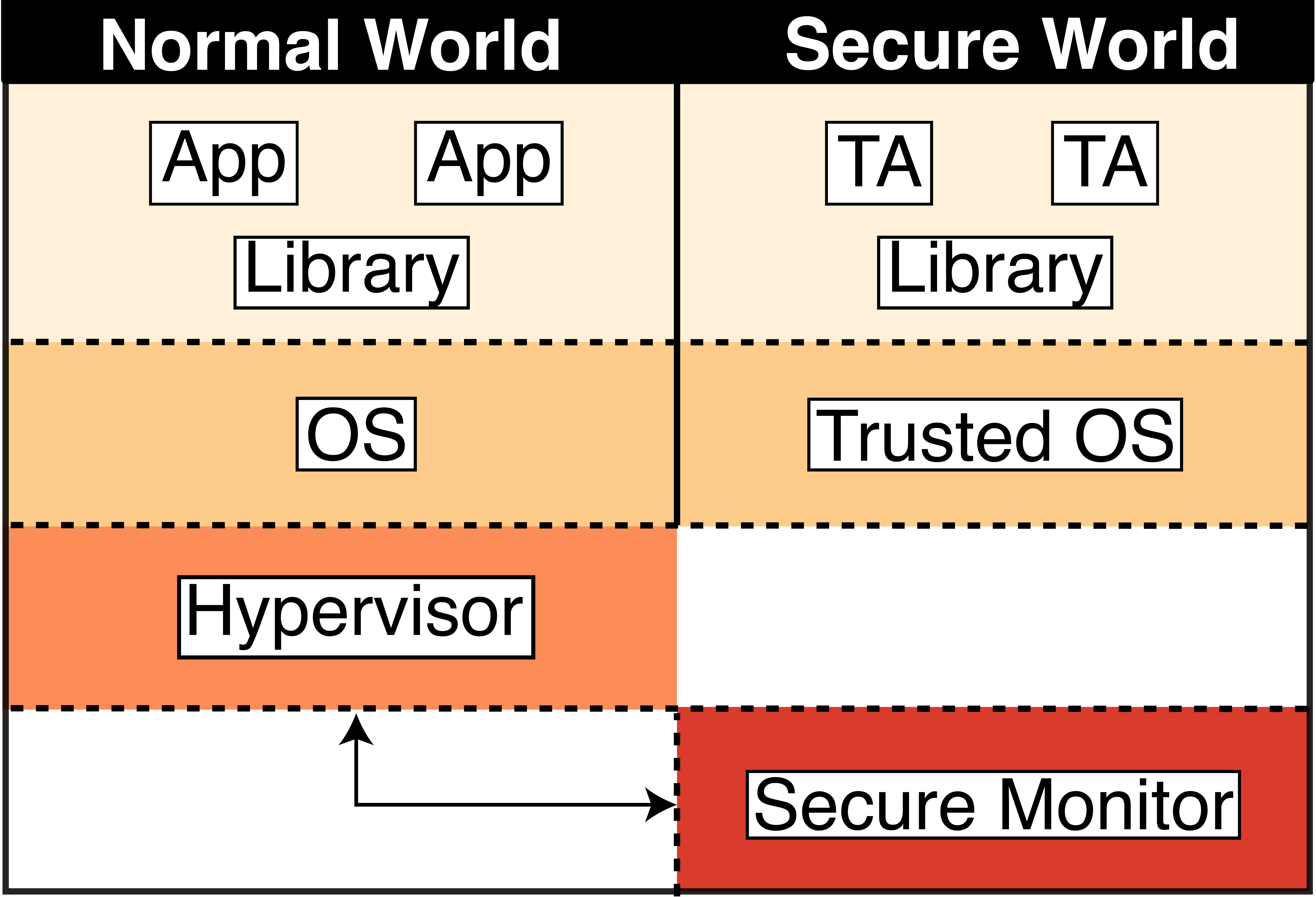}
        \caption{Arm TrustZone}
        \label{fig:trustzone}
    \end{subfigure}
    \hfill
    \begin{subfigure}[b]{0.3\textwidth}
        \centering
        \includegraphics[width=\textwidth]{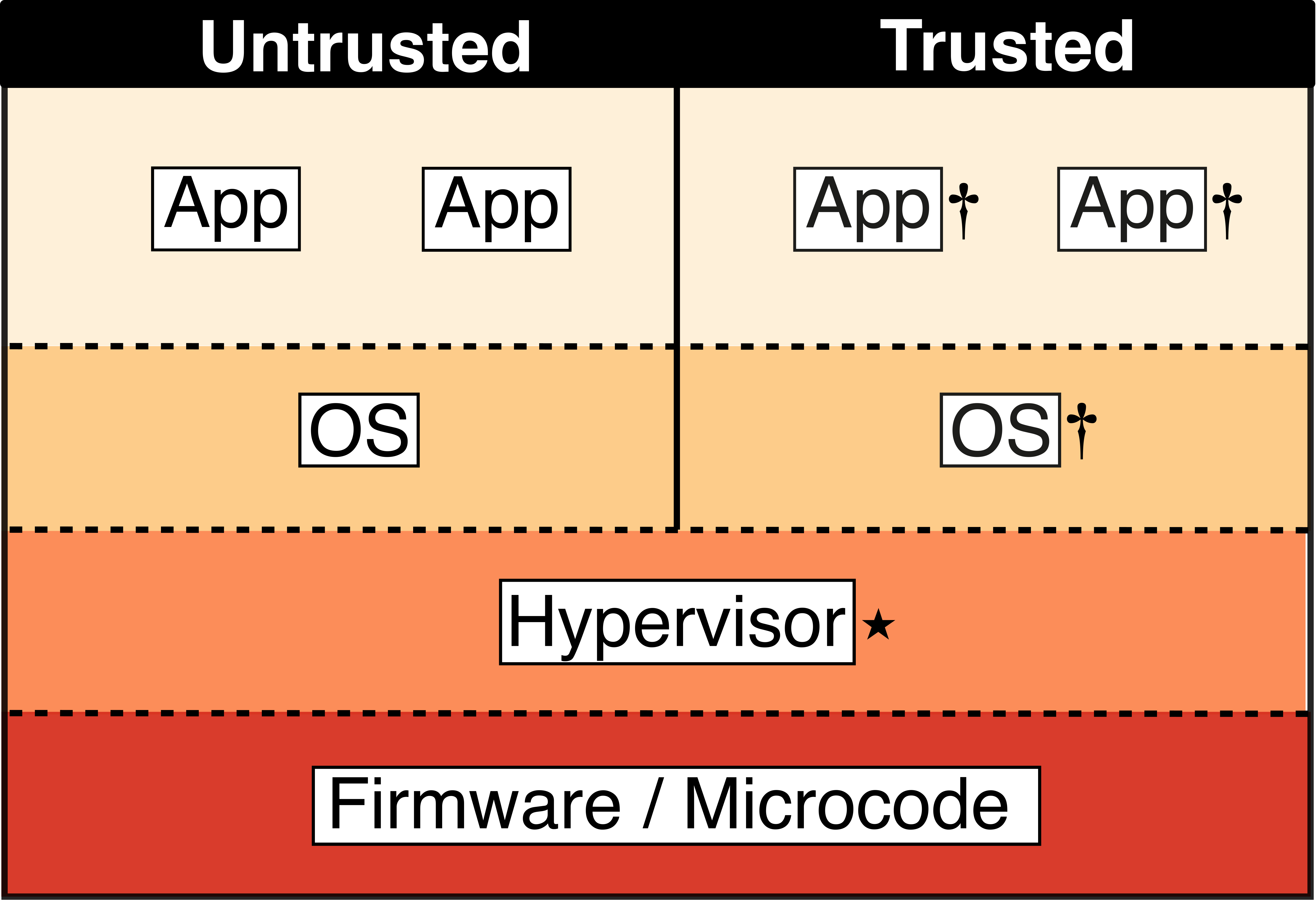}
        \caption{AMD SEV}
        \label{fig:sev}
    \end{subfigure}
    \vspace{6pt}
    \caption{High-level description of major TEE architectures (\textdagger\ indicates attested components, \textborn\ is untrusted for SEV-SNP).}
    \label{fig:tees}
\end{figure*}

Intel \emph{Software Guard Extensions} (SGX)~\cite{cryptoeprint:2016:086} introduced TEEs for mass-market processors in its Skylake architecture in 2015.
SGX is a set of instructions to create encrypted regions of memory, called \emph{enclaves}, protected in a special execution mode of the CPU.
Figure~\ref{fig:sgx} illustrates the high-level architecture of SGX.
A memory region is reserved at boot time for storing code and data of encrypted enclaves.
This memory area, called the \emph{Enclave Page Cache} (EPC), is inaccessible to other programs running on the same machine, including the operating system and the hypervisor.
The traffic between the CPU and the system memory remains confidential thanks to the \emph{Memory Encryption Engine} (MEE).
The EPC also stores verification codes to ensure that the RAM corresponding to the EPC was not modified by any software external to the enclave.

A trusted application executing in an enclave may establish a local attestation with another enclave running on the same hardware.
Reports are structures that are created and signed by the \texttt{EREPORT} instruction.
Reports contain identities, attributes (\ie modes and other properties), the trustworthiness of the \emph{Trusted Computing Base} (TCB is the amount of hardware and software that needs to be trusted), additional information for the target enclave and a \emph{Message Authentication Code} (MAC).
This principle is further extended by Intel thanks to the usage of the built-in \emph{quoting enclave}.
A trusted application can hence receive a cryptographically signed evidence, \ie, a \emph{quote}, which may be enhanced by additional information, such as a public key (\eg, used for establishing a communication channel).
The quote binds a genuine Intel SGX processor with the measurement of the application when loaded into the enclave.
This quote can then be forwarded to a relying party and be verified remotely using the Intel attestation service~\cite{anati2013innovative,brickell2007enhanced} or a dedicated public key infrastructure~\cite{scarlata2018supporting}.
Intel designed their remote attestation protocol based on the SIGMA protocol~\cite{10.1007/978-3-540-45146-4_24} and extended it to the \emph{Enhanced Privacy ID} (EPID).
While the quoting enclave is closed-source and the microcode~\cite{intel2021xucode} of Intel SGX are not disclosed, recent work analysed the TEE and its attestation mechanism formally~\cite{10.1145/3133956.3134098,9217791}.
The other components of SGX (\ie kernel driver and SDK) are open source.
MAGE~\cite{chen2020mage} further extended the remote attestation scheme of Intel SGX by offering mutual attestation for a group of enclaves without trusted third parties.

Unlike local attestation, remote attestation requires an asymmetric-key scheme, which is made possible by the quoting enclave.
The quoting enclave is a special enclave that has access to the device-specific private key through the \texttt{EGETKEY} instruction.
First, enclaves do a local attestation with the quoting enclave.
The quoting enclave replaces the MAC after verification by a signature created with the private device key.
The EPID scheme does not identify unique entities, but rather a group of signers.
Each signer belongs to a group, and the verifier checks the group's public key.
Quotes are signed by the EPID key, which is bound to the firmware version of the processor~\cite{anati2013innovative}.
The quoting enclave manages the EPID key and has exclusive access to it.

\begin{figure}
    \centering
    \includegraphics[width=\columnwidth]{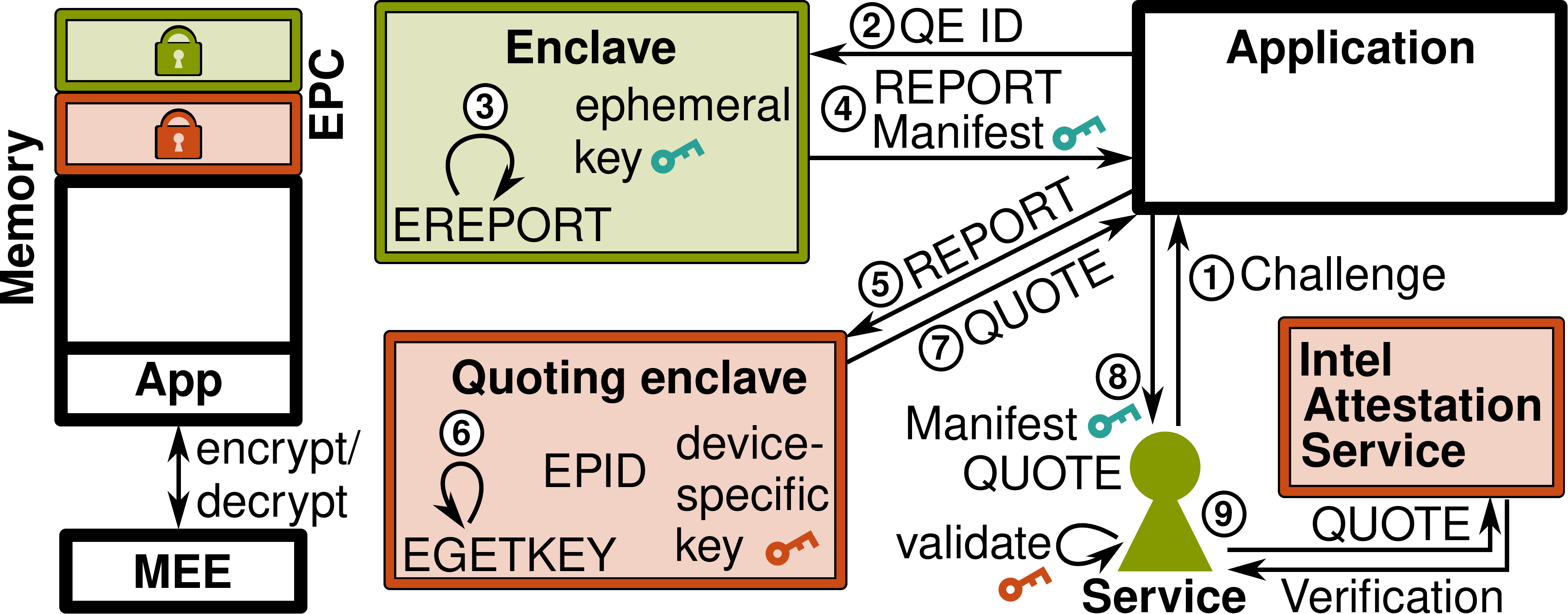}
    \caption{The remote attestation flow of Intel SGX.\label{fig:sgxra}}
  \end{figure}

In a remote attestation scenario, a service (\ie verifier) submits a challenge to the untrusted application with a nonce (Fig.\labelcref{fig:sgxra}-\ding{192}).
Together with the identity of the quoting enclave, the challenge is forwarded to the application enclave (Fig.\labelcref{fig:sgxra}-\ding{193}).
The application enclave (\ie attester) prepares a response to the challenge by creating a manifest (\ie a set of claims) and a public key (Fig.\labelcref{fig:sgxra}-\ding{194}), that is used to send back confidential information to the application enclave.
The manifest hash is used as auxiliary data in the report for the local attestation with the quoting enclave.
After verifying the report (Fig.\labelcref{fig:sgxra}-\ding{197}), the quoting enclave replaces the MAC with the signature from the EPID key and returns the quote (\ie evidence) to the application (Fig.\labelcref{fig:sgxra}-\ding{198}) which sends it back to the service (Fig.\labelcref{fig:sgxra}-\ding{199}).
The service verifies the signature of the quote (Fig.\labelcref{fig:sgxra}-\ding{200}) using either the EPID public key and revocation information or an attestation verification service~\cite{anati2013innovative}.
Finally, the service ensures the integrity of the manifest by verifying the response to the challenge.
\emph{Data Center Attestation Primitives} (DCAP)~\cite{intel2018dcap} is an alternative solution to EPID that enables third-party attestation for SGX of server-grade processors.

\subsection{Arm TrustZone architectures}

Arm TrustZone~\cite{pinto2019demystifying} provides the hardware elements to establish a single TEE per system.
Figure \ref{fig:trustzone} illustrates the high-level architecture of TrustZone.
Broadly adopted by commodity devices (including mobile devices, IoT edge nodes, \etc), TrustZone splits the processor into two states: the secure world (TEE) and the normal world (untrusted environment).
A secure monitor instruction (\ie the SMC) is switching between worlds, and each world operates with their own user and kernel spaces.
The trusted world uses a trusted operating system (such as OP-TEE~\cite{optee}) and runs \emph{Trusted Applications} (TAs) as isolated processes.
The normal world uses a traditional operating system such as Linux.

Despite the commercial success of TrustZone, it lacks attestation mechanisms, preventing relying parties from validating and trusting the state of a TrustZone TEE remotely.
Many protocols have been proposed for Arm TrustZone one-way remote attestation~\cite{10.1145/3319535.3363205,10.1145/2742647.2742676}, as well as for mutual remote attestation~\cite{ahn2020design,10.1145/3098954.3098971}, extending the capabilities of built-in hardware.
These protocols require the availability of extra hardware with a root of trust in the secure world, a secure source of randomness for cryptographic operations, and a secure boot mechanism.
Indeed, devices lacking built-in attestation mechanisms may rely on a secret fused in the die as a root of trust to derive private cryptographic materials (\eg a private key for evidence issuance).
Secure boot can measure the integrity of individual boot stages on devices and prevent tampered systems from being booted.
As a result, remote parties can verify issued evidences in the TEE and ensure the trustworthiness of the attesters.

In the following, we describe the remote attestation mechanism of Shepherd et al.~\cite{10.1145/3098954.3098971} as a study case.
This solution establishes mutually trusted channels for bi-directional attestation, based on a \emph{Trusted Measurer} (TM), which is a software component located in the trusted world and authenticated by the TEE's secure boot, to generate evidences based on the OS and TA states (\ie a set of claims).
A private key is provisioned and sealed in the TEE's secure storage and used by the TM to sign evidences, similarly to a firmware TPM~\cite{197213}.

Using a dedicated protocol for remote attestation, the bi-directional attestation is accomplished in three rounds.
First, the attester sends a handshake request to the verifier containing the identity of both parties and the cryptographic materials to initiate a key establishment.
Second, the verifier answers to the handshake by including similar information (\ie both identifies and cryptographic materials), as well as a signed evidence of the verifier's TEE, based on the computed common secret (\ie using Diffie-Hellman).
Finally, the attester sends back a signed evidence of the attester's TEE, based on the same common secret.
Once both parties assert that the evidences are genuine, they can derive common secrets to establish a trusted channel of communication.

\subsection{AMD SEV}

\begin{figure}[!t]
  \centering
  \includegraphics[width=\columnwidth]{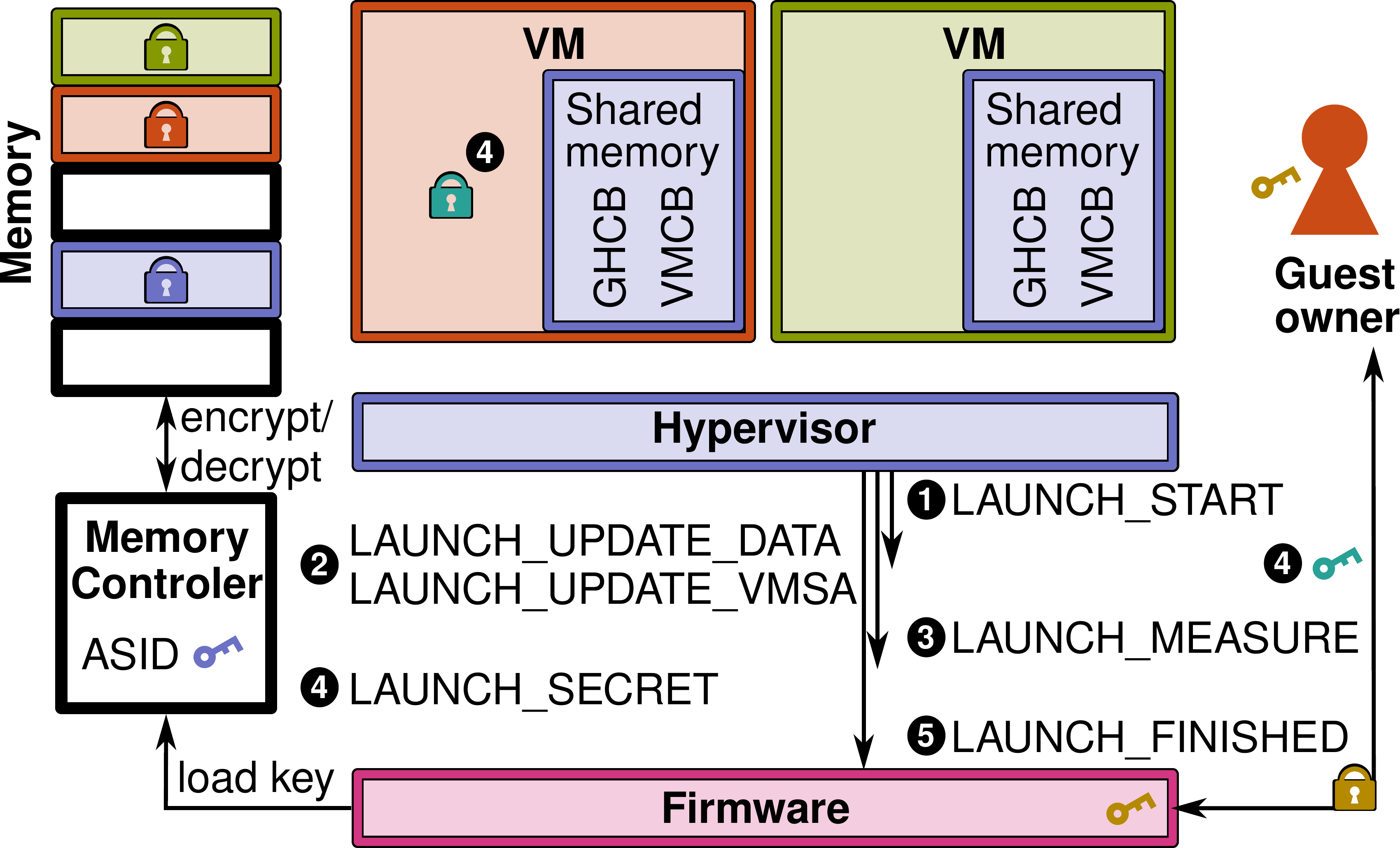}
  \caption{The remote attestation flow of AMD SEV.\label{fig:sevra}}
\end{figure}

AMD \emph{Secure Encrypted Virtualization} (SEV)~\cite{amd-sev} allows isolating virtualised environments (\eg containers and virtual machines) from trusted hypervisors.
Figure \ref{fig:sev} illustrates the high-level architecture of SEV.
SEV uses an embedded hardware AES engine, which seamlessly relies on multiple keys to encrypt memory. It exploits a closed Arm Cortex-v5 processor as a secure co-processor, used to generate cryptographic materials kept in the CPU.
Each virtual machine and hypervisor is assigned a particular key and tagged with an \emph{Address Space Identifier} (ASID), preventing cross-TEE attacks.
The tag restricts the code and data usage to the owner with the same ASID and protects from any unauthorised usage inside the processor.
Code and data are protected by AES encryption with a 128-bit key based on the tag outside the processor package.
SEV-ES (\emph{SEV Encrypted State})~\cite{kaplan2017seves} is a successor to SEV where register states are encrypted, and the guest operating system needs to grant the hypervisor access to specific guest registers, fixing an error in SEV that could leak sensitive information during interrupts from guests to the hypervisor through registers~\cite{hetzelt2017sev}.

Register states are stored with SEV-ES for each virtual machine in a \emph{Virtual Machine Control Block} (VMCB) that is divided into an unencrypted control area and an encrypted \emph{Virtual Machine Save Area} (VMSA).
The hypervisor manages the control area to indicate event and interrupt handling, while VMSA contains register states.
Integrity protection ensures that encrypted register values in the VMSA cannot be modified without being noticed and that virtual machines resume with the same state.
Requesting services from the hypervisor due to interrupts in virtual machines are communicated over the \emph{Guest Hypervisor Communication Block} (GHCB) that is accessible through shared memory.
Hypervisors do not need to be trusted with SEV-ES because they no longer have access to guest registers.
\emph{SEV Secure Nested Paging} (SNP)~\cite{sev2020strengthening} were proposed to prevent rollback attacks~\cite{buhren2019insecure} allowing a malicious cloud provider with physical access to SEV machines to easily install malicious firmware and be able to read in clear the (otherwise protected) system.

At its core, SEV leverages a \emph{Chip Endorsement Key} (CEK), a secret fused in the die of the processor and issued by AMD for its attestation mechanism.
The three editions of SEV may start the virtual machines from an unencrypted state, similarly to SGX enclaves.
In such cases, the secrets and confidential data must then be provisioned using remote attestations queries.
The AMD secure processor cryptographically measures the content of the virtual machine into a launch digest (\ie claim).
In addition, AMD-SNP measures the metadata associated with memory pages, ensuring the digest also considers the layout of the initial guest memory.
While SEV and SEV-ES only support remote attestation during the launch of the guest operating system, SEV-SNP supports a more flexible model.
That latter bootstraps private communication keys, enabling the guest virtual machine to request attestation reports (\ie evidence) at any time and obtain cryptographic materials for data sealing, \ie, storing data securely at rest.

SEV uses six launch commands for hypervisors to prepare encrypted memory before enabling SEV for virtual machines.
The \texttt{LAUNCH\_START} command (Fig.\labelcref{fig:sevra}-\ding{202}) creates a guest context in the firmware with the public key of the guest owner provided by the hypervisor.
As the hypervisor is loading the virtual machine into memory, \texttt{LAUNCH\_UPDATE\_DATA} commands (Fig.\labelcref{fig:sevra}-\ding{203}) are called to encrypt the memory and calculate measurements.
The hypervisor initialises the VMSA inside the VMCB with the \texttt{LAUNCH\_UPDATE\_VMSA} command, which is only available if SEV-ES is enabled.
When the virtual machine is loaded, and the VMSA is initialised, the hypervisor calls the \texttt{LAUNCH\_MEASURE} command (Fig.\labelcref{fig:sevra}-\ding{204}), which produces a measurement of the encrypted virtual machine.
The SEV firmware provides guest owners with the measurement containing a signature of the state of their virtual machine to prove that it is in the expected state.
The guest owner verifies that the virtual machine launched correctly and has not been interfered with before provisioning any sensitive data to the virtual machine.
Sensitive data, such as image decryption keys, is provisioned through the \texttt{LAUNCH\_SECRET} command (Fig.\labelcref{fig:sevra}-\ding{205}) after which the hypervisor calls the \texttt{LAUNCH\_FINISHED} command (Fig.\labelcref{fig:sevra}-\ding{206}) to indicate that the virtual machine can be executed.

Software development is eased, as AMD SEV protects the whole virtual machine, which comprises the operating system, compared to the SGX paradigm where the applications must be split in an untrusted and trusted part.
Nonetheless, this approach increases the attack surface of the secure environment since the TCB is enlarged.
The guest operating system must also support SEV, cannot access host devices (PCI passthrough), and the first edition of SEV (called \emph{vanilla} in Table \ref{tab:features}) is limited to 16 virtual machines.

\subsection{RISC-V architectures}
Several TEE designs were proposed for RISC-V based on its physical memory protection (PMP) instructions, some even including support for remote attestation. In the following, we describe the designs we deemed more important for the scope of our study.

Keystone~\cite{10.1145/3342195.3387532} is a modular framework that provides the building blocks to create trusted execution environments.
Keystone implements a secure monitor at machine mode (M-mode) and relies on the RISC-V PMP instructions, without requiring any hardware change.
Users can select their own set of security primitives, \eg, memory encryption, dynamic memory management and cache partitioning.
Each trusted application executes in user mode (U-mode) and embeds a runtime that executes in supervisor mode (S-mode).
The runtime decouples the infrastructure aspect of the TEE (\eg, memory management, scheduling) from the security aspect handled by the secure monitor.
Keystone utilises a secure boot mechanism that measures the secure monitor image, generates an attestation key and sign them using a hardware-visible secret (\ie root of trust).
The secure monitor exposes a \emph{Supervisor System Interface} (SBI) for the enclaves to communicate.
A subset of the SBI is dedicated to issue evidences signed by provisioned keys (\ie endorsed by the verifier), based on the measurement of the secure monitor, the runtime and the enclave's application.
Arbitrary data can be attached to the evidence, enabling an attester to create a secure communication channel with a verifier using standard protocols.
When a remote attestation request occurs, the remote party (\ie verifier) sends a challenge to the trusted application.
The response contains the evidence with the public session key of the attester.
Finally, the evidence is verified based on the public signature and the measurements of components (\ie claims).
While Keystone does not describe in-depth the protocol, the authors provide a case study of remote attestation.

Sanctum~\cite{197162} has been the first proposition with support for attesting trusted applications.
It offers similar promises to Intel's SGX by providing provable and robust software isolation, running in enclaves.
The authors replaced Intel's opaque microcode with two open-source components: the \emph{measurement root} (\texttt{mroot}) and a secure monitor as a means to provide verifiable protection.
A remote attestation protocol is proposed with a design for deriving trust from a root of trust.
Upon booting the system, \texttt{mroot} generates the necessary keys and hands off to the secure monitor.
Similarly to SGX, Sanctum owns a dedicated signing enclave, that receives a derived private key from the secure monitor to generate evidences.
The remote attestation protocol requires the attester to establish a session key with the verifier.
Afterwards, a regular enclave can request an evidence to the signing enclave based on multiple claims, such as the hash of the code of the requesting enclave and some information coming from the key exchange messages.
This evidence is then forwarded to the verifier by the secure channel previously established for examination.
This work has been further extended to establish a secure boot mechanism and an alternative method for remote attestation by deriving a cryptographic identity from manufacturing variation using a PUF, which is useful when a root of trust is not present~\cite{8429295}.

TIMBER-V~\cite{Weiser2019TIMBERVTM} achieved the isolation of execution on small embedded processors thanks to hardware-assisted memory tagging.
Tagged memory transparently associates blocks of memory with additional metadata.
Unlike Sanctum, they aim to bring enclaves to smaller RISC-V with limited physical memory.
Similarly to TrustZone, user and supervisor modes are split into secure and normal worlds.
The secure supervisor mode runs a trust manager, called \emph{TagRoot}, which manages the tagging of the memory.
The secure user mode improves the model of TrustZone, as it can handle multiple isolated enclaves.
They combine tagged memory with an MPU to support an arbitrary number of processes.
The trust manager exposes an API for the enclaves to retrieve an evidence, based on a given enclave identity, a secret platform key (\ie root of trust), and an arbitrary identifier provided by the trusted application.
The remote attestation protocol is twofold: the remote party (\ie verifier) sends a challenge to the trusted application (\ie attester).
Next, the challenge is forwarded to the trust manager as an identifier to issue an evidence, which is authenticated using a MAC.
The usage of symmetric cryptography is unusual in remote attestation because the verifier requires to own the secret key to verify the evidence.
The authors added that TIMBER-V could be extended to leverage public key cryptography for remote attestation.

LIRA-V~\cite{9474324} drafted a mutual remote attestation for constrained edge devices.
While this solution does not enable arbitrary code execution in a TEE, it introduces a comprehensive remote attestation mechanism.
The proposed protocol relies exclusively on machine mode (M-mode) or machine and user mode (M-mode and U-mode).
Claims are computed on parts of the device physical memory regions by a program stored in the ROM.
LIRA-V's mutual attestation is similar to the protocol illustrated in TrustZone-A, and requires provisioned keys as a root of trust.
The first device (\ie verifier) sends a challenge with a public session key.
Next, the second device (\ie attester) answers with a challenge and public session key, as well as an evidence bound to that device and encrypted using the established shared session key.
Finally, if the first device asserts the evidence, it becomes the attester and issues an evidence to be sent to the second device, which becomes the verifier.

We omitted some other emerging TEEs leveraging RISC-V as they lack a remote attestation mechanism.
For instance, SiFive, a provider of commercial RISC-V processors, proposes Hex-Five MultiZone~\cite{garlati2020clean}, a zero-trust computing architecture enabling the isolation of software, called \emph{zones}.
The multi zones kernel ensures the sane state of the system using secure boot and PMP and runs unmodified applications by trapping and emulating functionality for privileged instructions.
HECTOR-V~\cite{10.1145/3433210.3453112} is a design for developing hardened TEEs with a reduced TCB. Thanks to a tight coupling of the TEE and the SoC, the authors provide runtime and peripherals services directly from the hardware and leverage a dedicated processor and a hardware-based security monitor, which ensure the isolation and the control-flow integrity of the trusted applications, called \emph{trustlets}.
Finally, Lindemer et al.~\cite{lindemer2020real} enable simultaneous thread isolation and TEE separation on devices with a flat address space (\ie without an MMU), thanks to a minor change in the PMP specification.
 \section{Conclusion}
\label{sec:conc}
This work compares state-of-the-art remote attestation schemes, which leverage hardware-assisted TEEs, helpful for deploying and running trusted applications from commodity devices to cloud providers.
TEE-based remote attestation has not yet been extensively studied and seems to remain an industrial challenge.

Our survey highlights four architectural extensions: Intel SGX, Arm TrustZone, AMD SEV, and upcoming RISC-V TEEs.
While SGX competes with SEV, the two pursue significantly different approaches.
The former provides a complete built-in remote attestation protocol for multiple, independent, trusted applications.
The latter is designed for virtualized environments, shielding VMs from untrusted hypervisors, and provides instructions to help the attestation of independent VMs.
Arm TrustZone and native RISC-V do not provide means for attesting software running in the trusted environment, relying on the community to develop software-based alternatives.
However, TrustZone-M supports a root of trust, helping to develop an adequately trusted implementation in software.
RISC-V extensions differ a lot, offering different combinations of software and hardware extensions, some of which support a root of trust and multiple trusted applications.

Whether provided by manufacturers or developed by third parties, remote attestation remains an essential part of the design of trusted computing solutions.
They are the foundation of trust for remote computing where the target environments are not fully trusted.
Current solutions widely differ in terms of maturity and security.
Whereas some TEEs are developed by leading processor companies and provide built-in attestation mechanisms, others still lack proper hardware attestation support and require software solutions instead.
Our study sheds some light on the limitations of state-of-the-art TEEs and identifies promising directions for future work.

\section*{Acknowledgments}
This publication incorporates results from the VEDLIoT project, which received funding from the European Union’s Horizon 2020 research and innovation programme under grant agreement No 957197. 
\printbibliography

\end{document}